# Valley polarization in MoS$_2$ monolayers by optical pumping

Hualing Zeng[†,1], Junfeng Dai[†,1,3], Wang Yao[1], Di Xiao[2], Xiaodong Cui[1,*]

1. Physics Department, The University of Hong Kong, Pokfulam road, Hong Kong, China

2. Materials Science and Technology Division, Oak Ridge National Laboratory, Oak Ridge, Tennessee, 37831, USA

3. Physics Department, South University of Science and Technology of China, Shenzhen, 518005, China

† The authors contributed equally to this work

* To whom correspondence should be addressed.  Email: xdcui@hku.hk

**A trend for future electronics is to utilize internal degrees of freedom of electron, in addition to its charge, for nonvolatile information processing. A paradigmatic example is spintronics utilizing the spin of electrons.[1,2] Degenerate valleys of energy bands well separated in momentum space constitute another discrete degrees of freedom for low-energy electrons with long relaxation time. This has led to the emergence of valleytronics, a conceptual electronics based on manipulating valley index, much in the same way as the spin index is used in spintronics applications.[3-6] As the first step, a controllable way to selectively fill or deplete valleys, thereby producing a valley polarization, is of crucial importance, and is the subject of growing theoretical and experimental efforts.[3-11] Here we report experimental evidences on selective occupation of the degenerate valleys by circularly polarized optical pumping in MoS$_2$ monolayer, an emerging multi-valley 2D semiconductor with remarkable optical and**

**transport properties. Over 30% valley polarization has been observed via the polarization resolved luminescence spectra on pristine MoS$_2$ monolayers.  These results demonstrate the viability of optical valley control in MoS$_2$ monolayers, which also form the basis for integrated valleytronics and spintronics applications on this platform with strong valley-spin coupling.**[12]

In many crystalline materials, it often happens that the conduction band minima and valence band maxima are located at degenerate and inequivalent valleys in momentum space. Because of the large valley separation in momentum space, the valley index can be regarded as a discrete degree of freedom for low-energy carriers, which is robust against smooth deformation and low-energy phonons because of the large valley separation in momentum space. To utilize the valley index as an information carrier, the crucial step is to identify a process in which the valley carriers respond differently to external stimuli. Recently, a simple scheme based on inversion symmetry breaking has been proposed to realize the manipulation of the valley index via electric, magnetic, and optical means.[10,11] In particular, it was shown that inversion symmetry breaking can lead to contrasted circular dichroism in different k-space regions which takes the extreme form of contrasting optical selection rules at the high symmetry points of the Brillouin zone.[11] This enables valley-dependent interplay of electrons with light of different circular polarizations, in analogy to spin dependent optical activities in semiconductors such as GaAs.

Here we report our observation of selective photoexcitation of the degenerate valleys by circularly polarized optical pumping in MoS$_2$ monolayer, an emerging multi-valley 2D semiconductor with remarkable optical and transport properties.[13, 14] Bulk $MoS_2$ follows a hexagonal crystal layered structure with a covalently bonded S-Mo-S hexagonal quisi-2D network packed by weak Van der Waals forces. Previous studies showed that with the decrease in thickness, $MoS_2$ crossovers from an indirect-gapped semiconductor at multi-layers to a direct band-gap one at monolayer thickness.[14, 15] The direct band gap is in the visible frequency range (~ 1.9 eV) ideal for optical applications, and both conduction and valence band edges are located at K (K') points of the 2D hexagonal Brillouin zone. In addition to the changes in electronic structure, $MoS_2$ thin film also has a structural change as sketched in figure 1.a. The inversion symmetry, which presents in the bulk and in thin films with even number of layers, is explicitly broken in thin films with odd number of layers giving rise to valley-contrasting optical selection rule:[11,12,16] the inter-band transitions in the vicinity of K(K') point couples exclusively to right (left)-handed circularly polarized light σ+(σ-), as sketched in Figure 1. The direct band-gap transition at the two degenerate valleys together with this valley-contrasting selection rules suggest that one can optically generate and detect valley polarizations in $MoS_2$ monolayer.

Our main result is summarized below. We find that in $MoS_2$ monolayers, the photoluminescence (PL) has the same helicity of circularly polarized component as the excitation laser, a signature of the optically pumped valley polarization. Below 90 K, a PL circular polarization of 32% is observed, which decays with temperature. The PL polarization shows no dependence on the in-plane magnetic field. The absence of the Hanle effect is a strong signature that the polarized PL is from polarization of valley rather than spin, since the former cannot be rotated by the magnetic field. Moreover, we find PL from MoS$_2$ bilayers is unpolarized under the same excitation condition, consistent with the presence of inversion symmetry in bilayers.

MoS$_2$ flakes representatively shown in figure 2.a are mechanically exfoliated with sticky tapes on SiO$_2$/silicon substrates in a manner similar to the technique of producing graphene. The monolayer, bilayer and multilayer flakes could be identified by two characteristic Raman modes: the in-plane vibrational $E_{2g}^1$ mode and the out-of-plane vibrational $A_{1g}$ mode around 400cm$^{-1}$.[13]  Following Ref [13], we label the sample thickness according to the frequency difference between $E_{2g}^1$ and $A_{1g}$ mode: Δω=19cm$^{-1}$ to monolayers and Δω=21cm$^{-1}$ to bilayers respectively (Figure 2.b). The photoluminescence spectra around 1.9eV corresponding to excitons from direct inter-band transition [17] is also used as a monolayer and bilayer indicator owing to the transition from indirect- to direct-gapped semiconductor, as illustrated in Ref.[14,15]. The polarization sensitive photoluminescence measurement is carried out with a confocal like microscopic set up. The collimated backscattering light passes through a broadband 1/4λ wave plate, a beam displacing prism which separates the light beam to two orthogonally polarized beams, a depolarizer and then is focused to two spots at the entrance slit of the monochromator equipped with CCD.  The polarization resolved spectrum could be obtained by analyzing the two branches of dispersion on the CCD.  Details could be found in the supplementary information.

Figure 2.c displays circularly polarized luminescence spectra peaked around 1.9eV with right- and left-handed circularly polarized excitation (HeNe laser, 1.96eV) at near resonant condition at T=10K.  The luminescence corresponds to direct inter-band transition

at K(K') valley.  The helicity of the luminescence exactly follows that of excitation light. Namely the right-handed circularly polarized excitation generates right-handed luminescence, and so does the left.  To characterize the circular component in the luminescence spectra, we define a polarization coefficient $P = \frac{\pm I(\sigma\pm)}{I_{tot}}$ where $I(\sigma\pm)$ is the intensity of the left(right)-handed circular component and $I_{tot}$ is the total light intensity.  For a perfect circularly polarized light, $P=1(\sigma+)$ or $-1(\sigma-)$.  The luminescence spectra display a symmetric polarization for excitation with opposite helicities:  $P=0.32$ under $\sigma+$ excitation and $P=-0.32$ under $\sigma-$ excitation.  These behaviors are fully expected in the mechanism of valley dependent optical selection rule. Besides the unpolarized background, there also exists a linearly polarized component and the linear polarization shifts by about $25^\circ$ between $\sigma+$ and $\sigma-$ excitation. If one switches the excitation light to a higher energy at 2.33eV, no polarization could be observed in the luminescence spectra. We note that the valley selection rule is valid in the vicinity of K(K') point,[11,12,16] whereas the optical transition with 2.33eV occurs far away from the K points in band dispersion.

In many semiconductor systems such as GaAs bulk and heterostructures the circular polarization of luminescence with circularly polarized excitation originates from electron (hole) spin polarization due to the spin dependent optical selection rule.[18,19]  This mechanism can be excluded here by examining the change of PL spectra in an in-plane magnetic field (Figure 3.a).  With a non-zero in-plane g-factor, spin polarization will precess about the in-plane magnetic field, and the time average of spin project along $z$-axis could be found at $S_z = \int_0^\tau S_{Z0} \exp(-t/\tau_S) \cos(g\mu_B Bt/\hbar)dt$ where $S_{Z0}$ is the initial spin along z axis, $\tau_S$ is the spin relaxation time, $\mu_B$ is Bohr magneton, and $\tau$ is the lifetime of the photoexcited carriers. Consequently the polarization of luminescence under continuous wave excitation should follow $P(B) \approx \frac{P(B=0)}{1+(g\mu_B B\tau_S/\hbar)^2}$ where $P(B=0)$ is the polarization of luminescence without magnetic field. This is the well known Hanle effect. If we assume the spin relaxation time $\tau_S$ around the same order of photocarrier lifetime $\tau \sim 10ps$,[20] then the polarization $P(B=0.65T)$ would drop to a few percent of $P(B=0)$. As shown in Figure 3.a, however, there is no visible difference between the PL polarization at zero field and in an in-plane field of $B=0.65T$. So spin polarization can not explain the polarized PL observed here. Instead, this magnetic field independent PL polarization is a fully expected consequence of the valley polarization through the valley dependent selection rule: Since the in-plane magnetic field does not couple to the valley index, the valley polarization can not be rotated by the magnetic field, hence no Hanle effect can be observed.

A further evidence to attribute the polarized PL in monolayers to valley dependent optical selection rule lies in the comparison with PL spectra from $MoS_2$ bilayers (Figure 3.b). The luminescence from bilayers is relatively weaker than that from monolayers and the peak is red-shifted presumably owing to environmental screening induced weaker exciton effect. The striking difference from the monolayers is that the circular polarization of luminescence from bilayers is negligible under the same conditions. This difference could be easily understood: the polarized PL is a consequence of the valley dependent optical selection rule arising from inversion symmetry breaking in $MoS_2$ monolayers which has $D_{3h}^1$ symmetry [11,12]. In contrast, MoS$_2$ bilayers are composed of two structurally identical monolayers stacking with hexagonal symmetry: the S atoms in one layer directly sit upon/down the Mo atoms in the other, and have $D_{6h}^4$ symmetry. Inversion symmetry is preserved in the bilayer unit cell, and consequently the valley dependent selection rule is not allowed in bilayers.[11] Circularly polarized PL is observed in bilayers with high excitation power, suggesting that the heating effects could induce structural anisotropy and consequently break the inversion symmetry in the bilayer.

Figure 4 displays the typical temperature dependence of the circular polarization of luminescence from monolayers under circular excitation. The circular polarization shows flat around 31% below 90K and then dramatically drops with temperature increase. The little temperature dependence of polarization at low temperature implies that the inter-valley scattering ( $K \leftrightarrow K'$ ) mainly results from scattering with grain boundaries and atomically sharp deformations. Since the sample is a natural mining product, abundant impurities and vacancies presumably provide inter-valley scattering centers and populate conduction electron at the energetically degenerated K and K' valley. The linearly polarized component of luminescence also suggests possible coherent mixing of the two valleys.

As the temperature increases above 90K, the circular and linear components of the luminescence spectra gradually decrease, which is a signature that phonons dominate in the valley-scattering at high temperature. The steady state PL polarization is inversely proportional to the valley scattering rate $\gamma_v$. We intuitively attribute the depolarization mechanism at T > 90K to scattering by acoustic phonons near the K points of the Brillouin zone which can supply the momentum change for inter-valley scattering. The valley scattering rate is then proportional to the population of these phonons: $\gamma_v \propto \exp(-E_K/k_B T)$, where $E_K$ is phonon energy near K points. The solid curve in figure 4 is a fit assuming $\exp(E_K/k_B T)$ dependence, from which we extract $E_K \sim 240 \text{cm}^{-1}$, consistent with the acoustic phonon energy near K point reported in the bulk and monolayer. [21,22]

In summary, we observed circularly polarized luminescence from $MoS_2$ monolayers under circularly polarized excitation. The circular polarization originates from the contrasting selection rules for optical transition at K and K' valleys. It provides us a viable tool to generate and detect valley polarization in $MoS_2$ monolayers.

Acknowledgement: We thank Bairen Zhu, Lu Xie and Dongmei Deng for technique assistance. The project was supported by HKU10/CRF/08, AoE/P-04/08, and HKU701810P.

# Figure Caption

Figure 1. (a) Schematic of $MoS_2$ monolayer structure (left) and bulk unit cell (right). It clearly shows the spatial inversion symmetry breaking in monolayer. (b) Schematic of the proposed valley dependent selection rules at K and K' points in crystal momentum space: a left(right)-handed circularly polarized light σ+(σ-) only couples to the band edge transition at K(K') points for the sake of angular moment conservation and time reversal symmetry.

Figure 2. (a) Representative optical image of $MoS_2$ monolayer, bilayer and thin film flakes. (b) Two characteristic Raman spectra from different $MoS_2$ flakes (monolayer, bilayer and thin film): the in-plane vibrational $E_{2g}^1$ mode and the out-of-plane vibrational $A_{1g}$ mode around 400cm$^{-1}$. (c) Polarization resolved luminescence spectra under circularly polarized excitation of a HeNe laser at 1.96eV at 10K. A circular polarization of P=32±2% (-32±2%) is observed along out-of-plane direction with right(left)-handed circular excitation.

Figure 3. (a) Circular components of luminescence spectra at zero magnetic field (black) and in-plane magnetic field of 0.65T (red). The two curves overlap within the apparatus's resolution. (b) Top: schematic of $MoS_2$ bilayer unit cell. Bottom: Circular components of luminescence spectra from $MoS_2$ bilayer (green) and monolayer (black) under circular excitation of 1.96eV at 10K: Negligible circular polarization is observed on $MoS_2$ bilayers.

Figure 4. The circular polarization P as a function of temperature. The fitting curve (red) assuming an inter-valley scattering proportional to phonon population gives a phonon energy around 240cm$^{-1}$.

Figure 1.

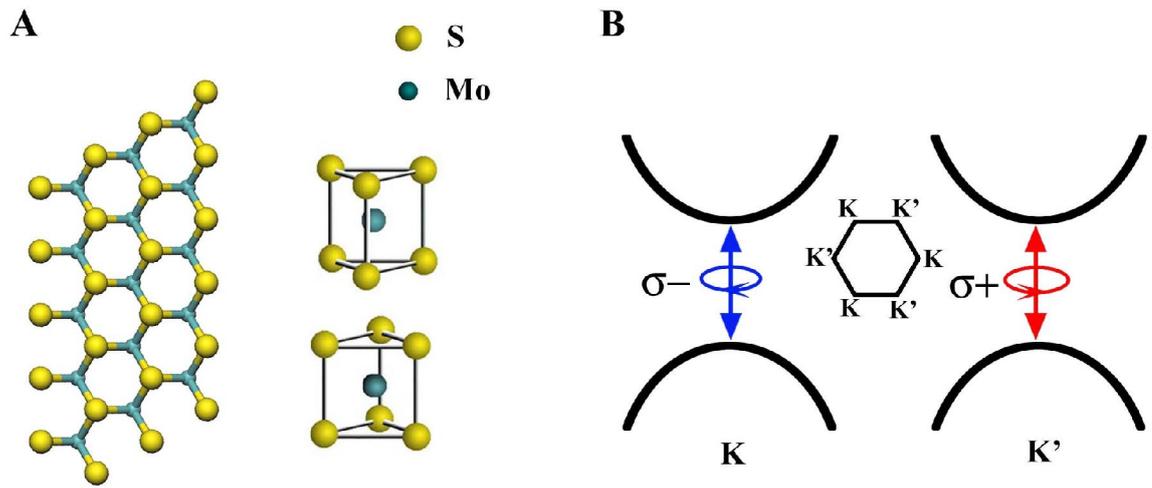

Figure 2

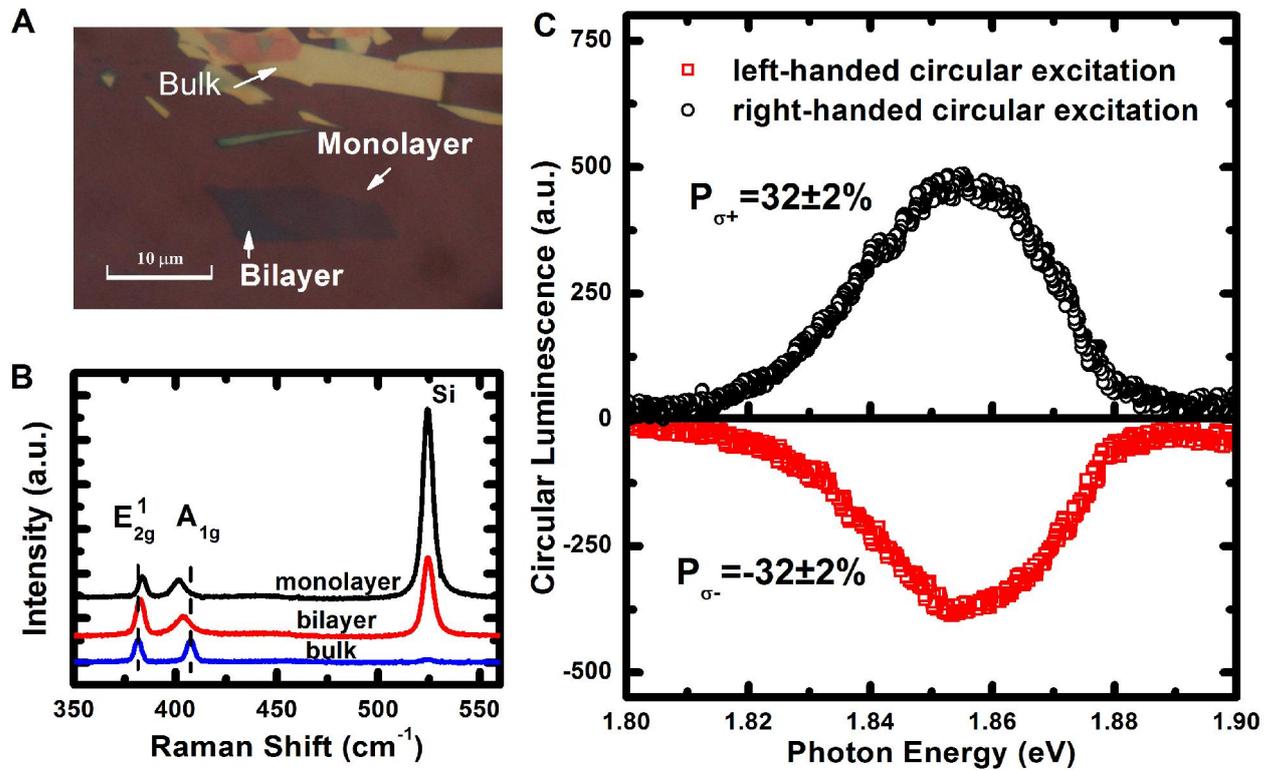

Figure 3

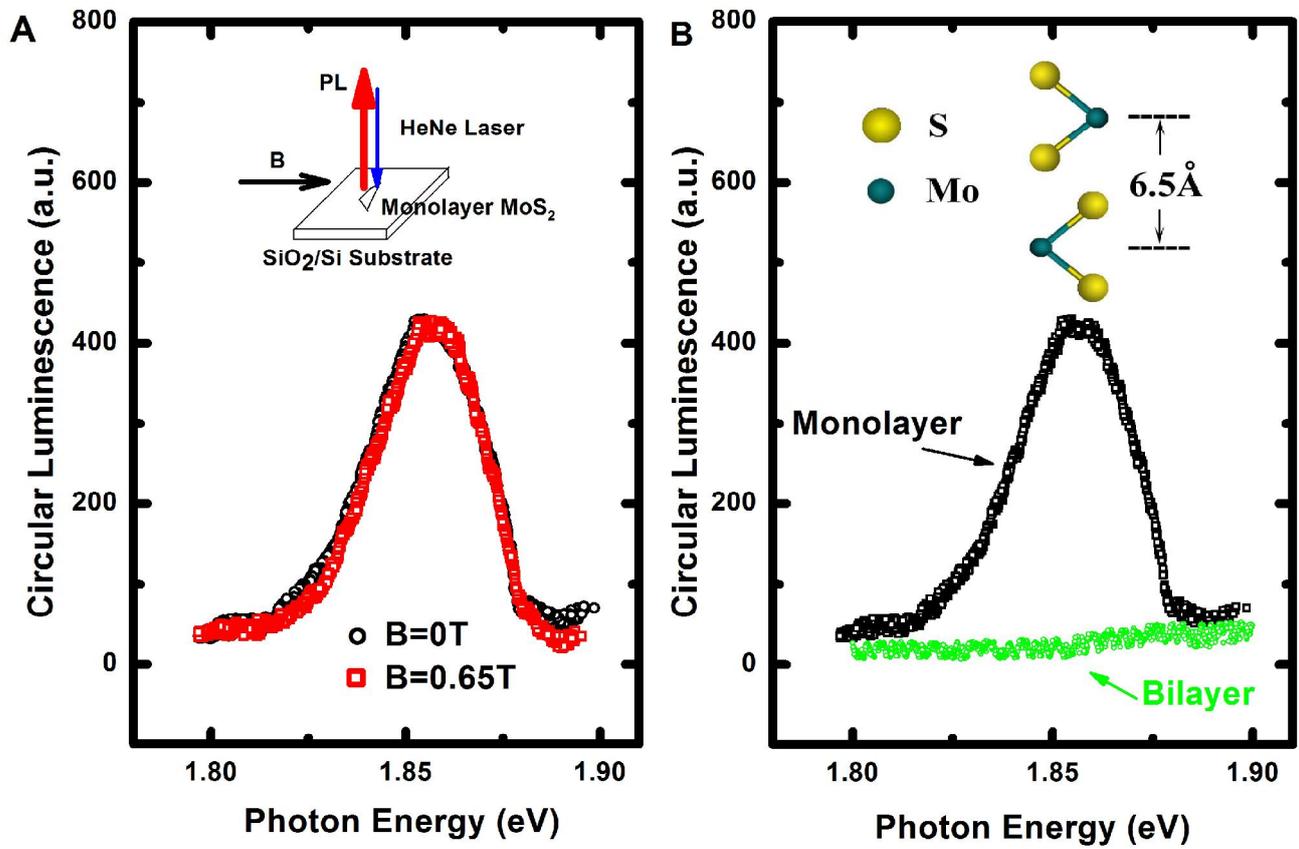

Figure 4

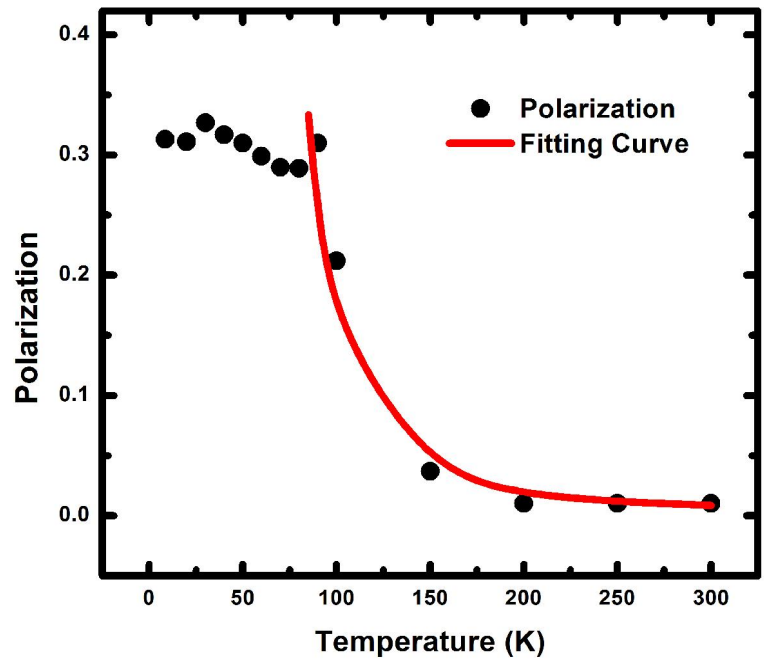